\documentclass[aps,twocolumn,showpacs,showkeys,amsmath,amssymb,prl]{revtex4} 

\usepackage{graphicx}   
\usepackage{dcolumn}    
\usepackage{bm}         
\usepackage{natbib}

\begin{document} 
  
\title{Bose-Hubbard phase diagram with arbitrary integer filling}

\author{Niklas Teichmann} \email{teichmann@theorie.physik.uni-oldenburg.de} 
\author{Dennis Hinrichs}
\author{Martin Holthaus}
\affiliation{Institut f\"ur Physik, Carl von Ossietzky Universit\"at, 
	D-26111 Oldenburg, Germany}
\author{Andr\'e Eckardt}
\affiliation{ICFO-Institut de Ci\`{e}ncies Fot\`{o}niques, 
	E-08860 Castelldefels (Barcelona), Spain}
\date{January 16, 2009}

\begin{abstract}
We study the transition from a Mott insulator to a superfluid in both the 
two- and the three-dimensional Bose-Hubbard model at zero temperature, 
employing the method of the effective potential. Converting Kato's perturbation
series into an algorithm capable of reaching high orders, we obtain accurate 
critical parameters for any integer filling factor. Our technique allows us 
to monitor both the approach to the mean-field limit by considering spatial 
dimensionalities $d > 3$, and to the quantum rotor limit of high filling, 
which refers to an array of Josephson junctions.      
\end{abstract}



\maketitle

The Bose-Hubbard model, describing interacting Bose particles moving on a
tight-binding lattice, has drawn much attention, especially after its 
experimental realization with ultracold bosonic atoms in optical potentials 
(see Ref.~\cite{BlochEtAl08} and references therein). This clean defectless 
setup, which allows for precise control of its parameters, has opened up new 
testing ground for quantum many-body physics. The pure Bose-Hubbard system 
reflects the competition between the potential energy due to the repulsive 
on-site interaction among the Bosons, which tends to suppress density 
fluctuations and to localize the particles, and the kinetic energy associated 
with tunneling processes between neighboring lattice sites, which try to
delocalize the particles and to reduce phase fluctuations. Denoting the  
on-site interaction energy of a pair of particles sitting at the same site 
by~$U$, and the hopping matrix element by~$J$, the model's grand canonical 
Hamiltonian is written in dimensionless form as~\cite{FisherEtAl89}
\begin{equation}
	H_{\rm BH} = \underbrace{\frac{1}{2} \sum_{j \phantom \rangle} 
	\hat{n}_j(\hat{n}_j-1) - \mu/U \sum_j \hat{n}_j}_{H_0} 
	\underbrace{- J/U \sum_{\langle j,k \rangle} \hat{a}_j^{\dagger} 
	\hat{a}_k^{\phantom \dagger}}_{H_{\rm tun}} \; ,	
\label{eq:Hamiltonian}
\end{equation} 
where indices label the sites of a $d$-dimensional lattice, which we take
as hypercubic, and the sum over $\langle j,k\rangle$ extends over nearest 
neighbors. As usual, $\hat{a}_j^{\dagger}$ and $\hat{a}_j^{\phantom \dagger}$ 
are the creation and annihilation operators for a Boson at site~$j$, and 
$\hat{n}_j = \hat{a}_j^{\dagger} \hat{a}_j^{\phantom \dagger}$ is the number 
operator at that site. The chemical potential $\mu$ here is site-independent. 
At zero temperature one finds a series of Mott phases at
sufficiently small values of $J/U$, characterized by a fixed filling of
an integer number of particles per site, depending on the value of $\mu/U$. 
A Mott state has zero compressibility, due to an energy gap separating the 
ground state from the particle and hole excitations, so that it costs energy 
to move a particle through the system. Upon incrasing the ratio $J/U$, the
competition between potential and kinetic energy leads to a quantum phase
transition: At the phase boundary $(J/U)_{\rm pb}$ the gap closes, so that 
particle delocalization becomes favorable, and the system Bose-condenses into 
a superfluid state for $d \ge 2$~\cite{FisherEtAl89}. In optical lattice 
experiments performed so far, this transition has been induced by varying 
the lattice depth~\cite{JakschEtAl98}, as in the pioneering work by Greiner 
{\em et al.\/}~\cite{GreinerEtAl02}, and by shaking the lattice periodically 
in time with slowly varying amplitude~\cite{EckardtEtAl05}, as done recently 
by Zenesini {\em et al.\/}~\cite{ZenesiniEtAl08} 

Despite the apparent simplicity of the Hamiltonian~(\ref{eq:Hamiltonian}), 
a precise calculation of its critical parameters for different 
dimensionalities~$d$ and filling factors~$g$ poses severe challenges, so 
that the determination of the phase diagram in the $J/U$--$\mu/U$-plane has 
become a major benchmark problem for computational many-body physics. Recent 
quantum Monte Carlo (QMC) simulations have yielded critical parameters with 
record accuracy for $g = 1$~\cite{CapogrossoSansone07,CapogrossoSansone08}. 
A previous strong-coupling expansion had led to reliable analytical results 
to third order in $J/U$~\cite{FreericksMonien96}, and later was extended to 
higher orders in one and two dimensions for $g = 1$ and 
$g = 2$~\cite{ElstnerMonien99}. Techniques using the density-matrix 
renormalization group (DMRG) allow one to treat fairly large 
systems in one dimension~\cite{KuehnerEtAl00,KollathEtAl04,KollathEtAl07}, 
but up to now have remained restricted to low filling. So far, accurate 
critical data for the three-dimensional (3D) system with experimentally 
relevant higher filling factors $g > 1$ have remained particularly hard 
to obtain.

In this contribution we show that a specific adaption of high-order many-body 
perturbation theory, based on Kato's formulation of the perturbation 
series~\cite{Kato49,Eckardt08} and using the concept of the order parameter,
enables one to investigate Bose-Hubbard systems with arbitrary integer filling 
factor. In principle, the technique is applicable to any type of lattice, in 
any dimension. We first briefly sketch the method, and present our results for 
both 2D and 3D lattices. We then numerically monitor the approach to the 
mean-field limit of high lattice dimension, and to the quantum rotor limit 
of high filling~\cite{Sachdev99,vanOostenEtAl03}, which describes a Josephson 
junction array~\cite{BruderEtAl05}.

Our starting point is the method of the effective 
potential~\cite{NegeleOrland98}, as considered recently by dos Santos and 
Pelster~\cite{SantosPelster08}. Adding source terms to the Bose-Hubbard 
Hamiltonian~(\ref{eq:Hamiltonian}) which attempt to add particles with 
uniform strength~$\chi$ to each site, or to remove them with strength~$\chi^*$
according to   
\begin{equation}
	\tilde{H}_{\rm BH}(\chi, \chi^*) =
	H_0 + H_{\rm tun} + \sum_j  \left(\chi^* \hat{a}_j^{\phantom \dagger} 
	+ \chi \hat{a}_j^{\dagger}\right) \; ,
\label{eq:BH_source}
\end{equation} 
then expanding the grand-canonical free energy
$F = \langle \tilde{H}_{\rm BH} \rangle$ at zero temperature into a power 
series in the hopping parameter~$J/U$ and the sources $\chi$, $\chi^*$, 
one has
\begin{equation}
	F(J/U,\chi, \chi^*) = 
	M \left(F_0(J/U) + \sum_n c_{2n}(J/U) |\chi|^{2n} \right)	 
\label{eq:free_energy}
\end{equation}
for a lattice of $M$ sites, with coefficients
\begin{equation}
	c_{2n}(J/U) = \sum_{\nu} \alpha_{2n}^{(\nu)} (J/U)^{\nu} \; .
\end{equation}
The order parameter $\psi$ now specifies the change of~$F$ in response to 
a variation of the sources, 
\begin{equation}
	\psi = \langle \hat{a}_j^{\phantom \dagger} \rangle 
	= \frac{1}{M}\frac{\partial F}{\partial \chi^*}
	\quad \text{and} \quad 
	\psi^* = \langle \hat{a}_j^{\dagger} \rangle 
	= \frac{1}{M}\frac{\partial F}{\partial \chi} \; ,
\label{eq:order_parms}
\end{equation} 
while the effective potential $\Gamma = F/M - \psi^* \chi - \psi \chi^*$ 
is the Legendre transform of~$F$, with $\psi$ and $\psi^*$ as independent 
variables. With the help of Eqs.~(\ref{eq:order_parms}) and 
(\ref{eq:free_energy}) one gets the familiar Landau form 
\begin{equation}
	\Gamma(J/U,\psi, \psi^*) = 
	F_0 - \frac{1}{c_2}|\psi|^2 + \frac{c_4}{c_2^4}|\psi|^4 + \ldots
	\; .
\label{eq:effpot}
\end{equation}
Since $\partial\Gamma/\partial \psi = -\chi^*$ and 
$\partial\Gamma/\partial \psi^* = -\chi$, and since the original Bose-Hubbard 
system is recovered by setting $\chi = \chi^* = 0$, the system adopts that
order parameter which minimizes~$\Gamma$. Unless $\mu/U$ is integer, one has 
$c_2 < 0$ for sufficiently small $J/U$, whereas $c_4 > 0$, so that one finds  
a Mott regime with $\psi = 0$. Upon increasing $J/U$, the system enters the
superfluid phase when $\psi$ acquires a nonzero value, indicating long-range
phase coherence. Hence, the phase boundary is determined by that $J/U$ for 
which the minimum of the expression~(\ref{eq:effpot}) starts to deviate from 
$|\psi|^2 = 0$, which occurs when the coefficient $-1/c_2$ of $|\psi|^2$ 
vanishes. In effect, one has to identify that scaled hopping strength~$J/U$ 
for which the susceptibility $c_2$ diverges; this divergence marks the quantum phase 
transition~\cite{SantosPelster08}. 

For computing $c_2$ we resort to Kato's formulation of the perturbation
series~\cite{Kato49,Eckardt08}, starting from the site-diagonal 
Hamiltonian~$H_0$. For integer filling factor~$g$, its ground state 
$|\mathbf{m}\rangle$ is a product of local Fock states with $g$ particles 
sitting at each site. In general, when the system is subjected to some 
perturbation~$V$, the $n$th-order correction to its energy is given by 
the trace~\cite{Kato49} 
\begin{eqnarray}	
	E_{|\mathbf{m}\rangle}^{(n)} 
	= {\rm tr} \left[ \sum_{ \{\alpha_\ell\} } 
	S^{\alpha_1} V S^{\alpha_2} V S^{\alpha_3} \ldots 
	S^{\alpha_n}VS^{\alpha_{n+1}}  \right] \; ,
\label{eq:Kato_energy}
\end{eqnarray}
where the sum runs over all possible sets of nonnegative integers 
$\alpha_\ell$ which obey $\sum_\ell \alpha_\ell = n-1$. The operators 
$S^{\alpha}$ are defined by 
\begin{equation}
S^{\alpha} = \left\{\begin{matrix} 
	-|\mathbf{m}\rangle \langle \mathbf{m}| & 
	\quad \text{for } \alpha = 0 \\
	\displaystyle
	\sum\limits_{i\neq \mathbf{m}} 
	\frac{|i \rangle \langle i|}{(E_{\mathbf{m}} - E_i)^{\alpha}} & 
	\quad \text{for } \alpha > 0               
	\end{matrix}\right. \; ,
\end{equation}
with $E_{\mathbf{m}}$ and $E_i$ denoting the unperturbed energies of the 
$H_0$-eigenstates $|\mathbf{m}\rangle$ and $|i\rangle$, respectively. This 
expression~(\ref{eq:Kato_energy}) can be understood as a sum over chains of 
processes mediated by the operators~$V$. Each process chain leads from the 
Mott-insulator state $|\mathbf{m}\rangle$ over different intermediate states 
$|i\rangle$ back to $|\mathbf{m}\rangle$. Such chains can be represented by 
abstract diagrams, with only connected diagrams contributing to the sum, as 
stated by the linked-cluster theorem~\cite{GelfandEtAl90}. Each diagram has 
a certain weight depending on the lattice's type and dimensionality. For 
example, diagrams for the energy correction due to tunneling consist merely 
of closed loops of individual tunneling processes. In contrast, for calculating
$c_2$ the augmented Hamiltonian~(\ref{eq:BH_source}) prompts us to set  
\begin{equation}
	V = - J/U \sum_{\langle j,k \rangle} 
	\hat{a}_j^{\dagger} \hat{a}_k^{\phantom \dagger}
	+ \sum_j \left( \chi^* \hat{a}_j^{\phantom \dagger}  
	+ \chi \hat{a}_j^{\dagger} \right) \; .
\end{equation} 
Because we are aiming at the coefficient of $|\chi|^2$ 
in Eq.~(\ref{eq:free_energy}), we only need to take into account terms 
containing exactly one creation and one annihilation process. This 
selection yields $c_2(J/U) = \sum_{\nu} \alpha_2^{(\nu)} (J/U)^{\nu}$ as a 
series in the tunneling parameter $J/U$. The only relevant third-order diagram 
thus consists of one creation of a Boson ($\bullet$), one tunneling process 
($\rightarrow$), and one annihilation ($\times$). The fourth-order diagrams 
then read
\begin{equation}
	\bullet \rightarrow \rightarrow \times \; , \qquad  
	\bullet \times \leftrightarrows \; ,
\label{eq:4_diagrams}
\end{equation} 
with the second diagram indicating chains for which creation and annihilation 
take place at the same site. The computational effort increases quickly with 
the order: For $\nu=8$, say, all permutations of up to ten different processes 
(8~$\rightarrow$, 1~$\bullet$, 1~$\times$) encoded in the diagrams have to 
be evaluated.

An instructive example for illustrating this scheme occurs in the limit of 
infinite lattice dimensionality~$d$. Here the diagrams containing ``back and forth''
tunneling processes (analogous to the second diagram~(\ref{eq:4_diagrams})) 
do not contribute to the sum, because they acquire vanishing weight for
$d \to \infty$. The remaining diagrams simply are 
\begin{equation}
	\bullet \times \; , \;  
	\bullet \rightarrow \times \; , \;
	\bullet \rightarrow \rightarrow \times \; , \; \ldots \; , \;
	\bullet \rightarrow \rightarrow \ldots \rightarrow \times \; .
\end{equation}
Being one-particle reducible, they factorize into their one-particle 
irreducible contributions~\cite{SantosPelster08,Kleinert06}:
\begin{eqnarray}
	\bullet\rightarrow \times &=& 
	(-1)^{\phantom 2} \left(\bullet \times \right)^2  
\nonumber\\
	\bullet \rightarrow \rightarrow \times &=& 
	(-1)^2 \left(\bullet \times \right)^3  
\\	\vdots & & \vdots			 
\nonumber\\
	\bullet ( \rightarrow )^\nu \times &=& 
	(-1)^\nu \left(\bullet \times \right)^{\nu+1}  
\nonumber \; .
\end{eqnarray}
For each tunneling process one has an additional factor $2d$, since 
there exist $2d$ directions on a $d$-dimensional rectangular lattice. 
The resulting  series for $c_2(J/U)$ is geometric, because 
$\alpha_2^{(\nu-1)}/\alpha_2^{(\nu)} = -1/(2d\alpha_2^{(0)}) $ is constant;
this ratio determines its radius of convergence and hence directly gives 
the phase boundary: 
\begin{equation}
	2d \, (J/U)_{\rm pb} = \frac{(g-\mu/U)(\mu/U-g+1)}{\mu/U+1} \; ,
\label{eq:mfboundary}
\end{equation} 
which is precisely the mean-field result~\cite{FisherEtAl89,Sachdev99}.

\begin{figure}
\includegraphics[scale=0.35,angle=-90]{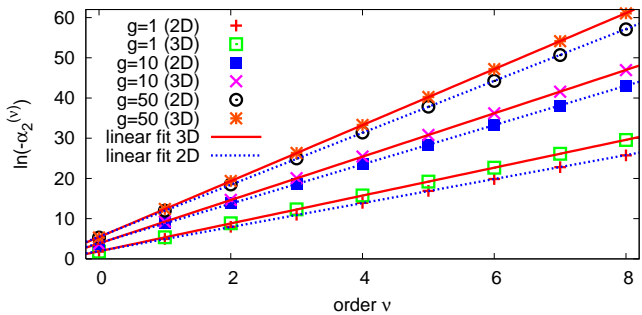}
\caption{Logarithm of the coefficients $-\alpha_2^{(\nu)}$ for filling 
	factors $g = 1, 10, 50$ in two and three  dimensions, with linear fits.
	The chemical potential is chosen as $\mu/U = g-0.5 $.
\label{fig:coeffs_c2}}	
\end{figure}

\begin{figure}
\includegraphics[scale=0.35,angle=-90]{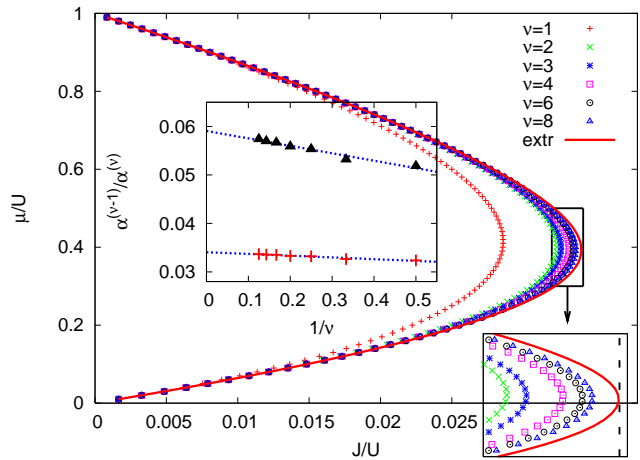}
\caption{Phase boundary for the 3D model with unit filling, as determined 
	from the ratios $\alpha^{(\nu-1)}/\alpha^{(\nu)}$ for finite 
	orders~$\nu$, together with the extrapolation to $\nu = \infty$ 
	(extr). The inset at the right bottom magnifies the tip 
	of the lobe, and demonstrates the convergence to the QMC 
	result~\cite{CapogrossoSansone07} (dashed vertical line). 
	The central inset illustrates the extrapolation of 
	$\alpha^{(\nu-1)}/\alpha^{(\nu)}$ to $(J/U)_{\rm c}$ for $d = 2$
	(upper data) and $d = 3$ (lower data). Observe that the data for 
	$d = 3$ fluctuate less.} 
\label{fig:extrapol_J_c}		
\end{figure}

We have devised an algorithm for efficiently generating and evaluating 
all diagrams up to some order for any lattice dimension~$d$. 
In two and three dimensions we obtain (negative) coefficients 
$\alpha_2^{(\nu)}$ which form almost perfect geometric series, as 
depicted in Fig.~\ref{fig:coeffs_c2} for $g = 1$, $10$, and $50$. If the 
ratio $\alpha_2^{(\nu-1)}/\alpha_2^{(\nu)}$ were constant, it would equal 
the phase boundary as in the example above. But since now this ratio 
changes slightly with the number~$\nu$ of tunneling processes taken into 
account, we carry out an extrapolation over $1/\nu$ by making a linear 
fit based on the orders $1$ to $8$ in $J/U$ ($3$ to $10$ in $V$), 
as illustrated by the central inset 
in Fig.~\ref{fig:extrapol_J_c}. Different selections of the orders employed 
(e.g., $2$ to $8$ in $J/U$) lead to very similar results, with an uncertainty 
of about 1\% in 3D, and 2\% in 2D. The main part of Fig.~\ref{fig:extrapol_J_c}
shows the phase boundary thus obtained for the 3D case at unit filling, 
together with some approximants for finite orders. The tip of the lobe 
corresponds to the critical parameter $(J/U)_{\rm c}$, for which QMC 
calculations have provided a highly accurate reference value: 
$(J/U)_{\rm c} = 0.03408(2)$ for $g = 1$~\cite{CapogrossoSansone07}. Our data 
match this value fairly well, as emphasized by the lower right inset.

\begin{figure}
\includegraphics[scale=0.3,angle=-90]{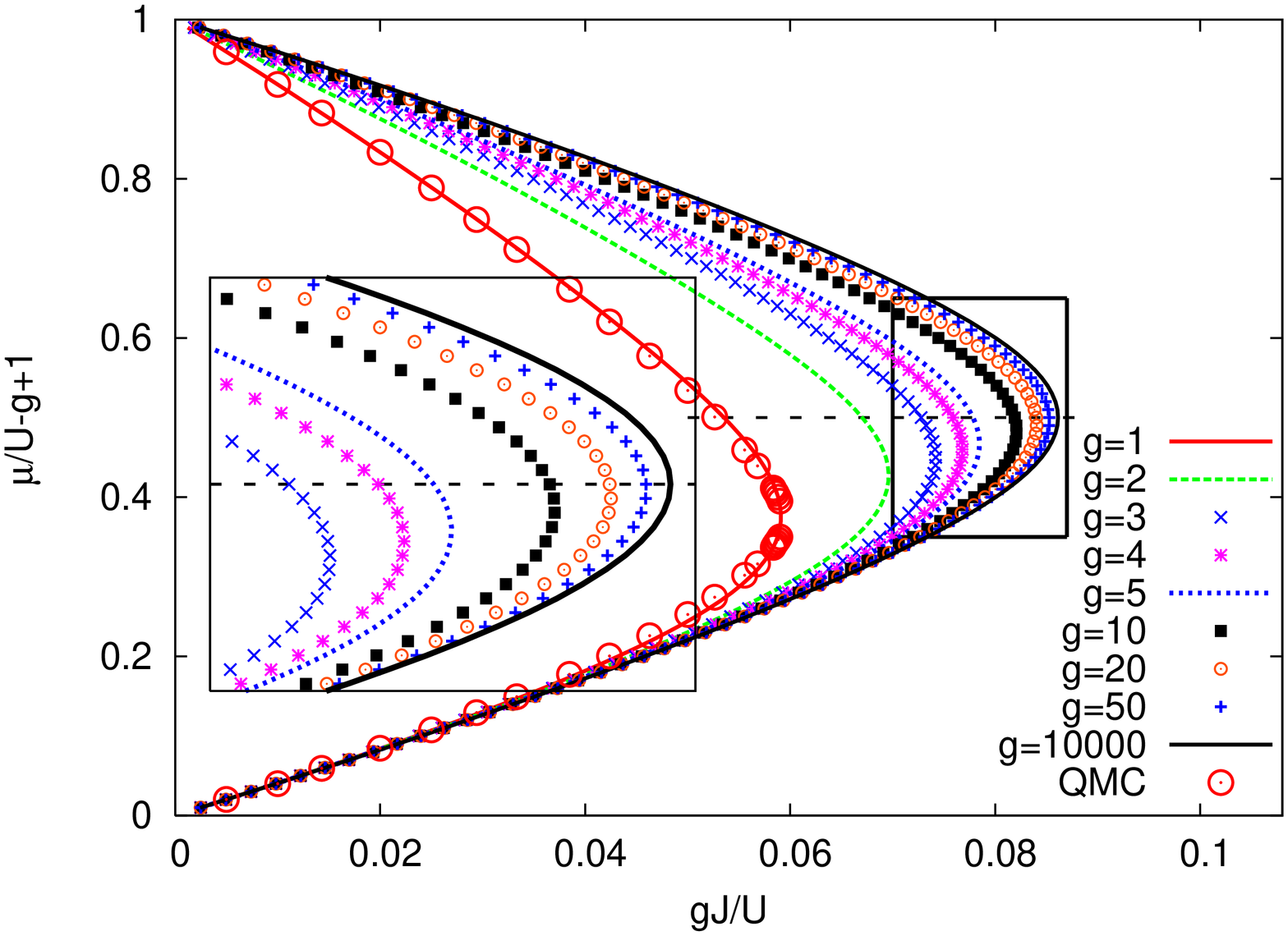}
\includegraphics[scale=0.3,angle=-90]{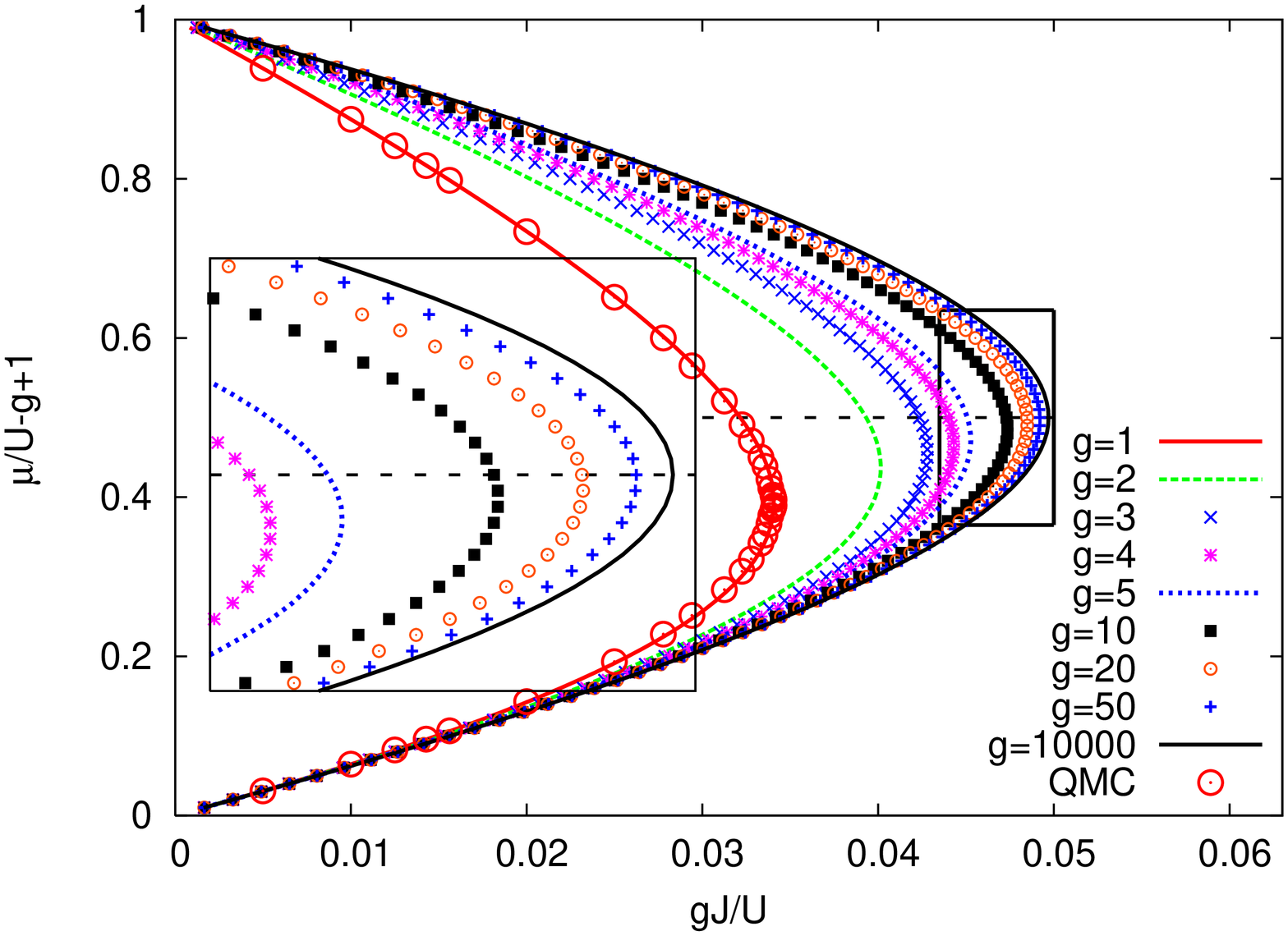}
\caption{Mott lobes for $d = 2$ (upper panel) and $d = 3$ (below) 
	with various~$g$. Dashed lines mark the quantum rotor limit 
	$(\mu/U)_c = g - 0.5$ of the critical chemical potential. 
	The lobes' tips are magnified in the inset, illustrating the 
	convergence of $g(J/U)_{\rm c}$. For unit filling, QMC 
	data~\cite{CapogrossoSansone07,CapogrossoSansone08} are included.} 
\label{fig:phase_dia}	
\end{figure}

Critical parameters obtained for higher filling~$g$ in two and three dimensions
are collected in Tab.~\ref{tab:J_c}. With increasing~$g$, the critical chemical
potential $(\mu/U)_{\rm c}$ approaches $g - 0.5$, due to the fact that there
is exact particle-hole symmetry for $g \to \infty$. Some corresponding Mott 
lobes are depicted in Fig.~\ref{fig:phase_dia}; for $g = 1$, 
QMC data~\cite{CapogrossoSansone07,CapogrossoSansone08} are included for 
comparison.

\begin{table}
\caption{Critical values $(\mu/U)_{\rm c}$ and $(J/U)_{\rm c}$ for various 
	filling factors~$g$. For locating the tip of the respective Mott lobe, 
	$\mu/U$ has been varied in steps of $0.001$. Relative errors of 
	$(J/U)_{\rm c}$ are less than 1\% for $d = 3$, and less than 2\% for 
	$d = 2$.}
\label{tab:J_c}
\begin{ruledtabular}
\begin{tabular}{r|r|r|r|r}
      & \multicolumn{2}{c|}{$d=2$} & \multicolumn{2}{c}{$d=3$} \\ \hline
      $g$ & $(\mu/U)_c$ & $(J/U)_c$ & $(\mu/U)_c$    & $(J/U)_c$   \\ \hline
        1 & 0.376        & 5.909E-002   &  0.393       & 3.407E-002       \\
        2 & 1.427        & 3.480E-002   &  1.437       & 2.007E-002       \\
        3 & 2.448        & 2.473E-002   &  2.455       & 1.427E-002       \\
        4 & 3.460        & 1.920E-002   &  3.465       & 1.108E-002       \\
        5 & 4.470        & 1.569E-002   &  4.472       & 9.055E-003       \\
       10 & 9.483        & 8.208E-003   &  9.485       & 4.736E-003       \\
       20 & 19.491       & 4.202E-003   &  19.492      & 2.425E-003       \\
       50 & 49.496       & 1.706E-003   &  49.497      & 9.842E-004       \\
      100 & 99.498       & 8.571E-004   &  99.498      & 4.946E-004       \\
     1000 & 999.50       & 8.609E-005   &  999.50      & 4.968E-005       \\
    10000 & 9999.50      & 8.613E-006   &  9999.50     & 4.970E-006
\end{tabular}
\end{ruledtabular}	
\end{table}

\begin{figure}
\includegraphics[scale=0.35,angle=-90]{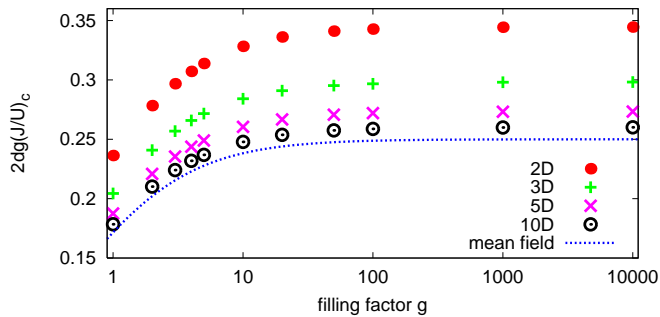}
\caption{Critical product $2dg (J/U)_{\rm c}$ for $d = 2$, $3$, $5$, and 
	$10$ vs.~$g$, together with the mean-field limit. Even for $d = 10$, 
	the large-$g$-limit still exceeds the mean-field prediction by 4\%.}
\label{fig:crit_hop}	
\end{figure}

Our technique permits us to reach higher dimensionalities $d > 3$, thus 
uncovering how the mean-field limit is approached, and high filling factors 
$g \gg 1$. In the latter regime, the phases at the individual sites become 
well defined, so that the Bose-Hubbard model reduces to a quantum rotor model 
containing a single parameter $gJ/U$, and describing a Josephson junction 
array~\cite{Sachdev99,BruderEtAl05}. Figure~\ref{fig:crit_hop} indeed reveals 
that the products $2dg(J/U)_{\rm c}$ remain almost constant when~$g$ exceeds 
$100$, with limiting values $0.345$ for $d=2$ and $0.299$ for $d=3$ falling 
significantly above the mean-field prediction of $1/4$, which follows from 
Eq.~(\ref{eq:mfboundary}). Even for $d=10$, the data still exceed the 
mean-field result by 4\%.

To conclude, diagrammatic many-body perturbation theory based on Kato's
series~(\ref{eq:Kato_energy}), though impractical to work out analytically
in high orders, becomes a powerful and accurate tool when turned into 
a numerically executable algorithm. The merit of this technique rests 
not only in the fact that it enables one to access regimes which could not 
be reached before, such as experimentally important filling factors 
$g > 1$~\cite{BlochEtAl08}, or the crossover to the quantum rotor dynamics 
depicted in Fig.~\ref{fig:crit_hop}, but also in its great flexibility. 
For instance, with appropriately constructed diagrams it also yields 
correlation functions. Thus, the applicability of this approach is by no 
means exhausted by the present calculation of the Bose-Hubbard phase diagram.

\begin{acknowledgments}
We thank F.~Gebhard and A.~Pelster for insightful discussions, 
and B.~Capogrosso-Sansone for providing the QMC 
data~\cite{CapogrossoSansone07,CapogrossoSansone08}.
Computer power was obtained from the GOLEM~I cluster of the Universit\"at
Oldenburg. N.~T.\ acknowledges a fellowship from the Studienstiftung des 
deutschen Volkes. A.~E.\ thanks M.\ Lewenstein for kind hospitality, and 
acknowledges a Feodor Lynen research grant from the Alexander von Humboldt 
foundation.
\end{acknowledgments}

\end{document}